\documentstyle[prl,aps,preprint,tighten]{revtex}

\begin{document}
\draft


\preprint{\vbox{\it Submitted to Phys.\ Rev.\ C --- Rapid Communications
\hfill\rm DOE/ER/40762--039\\
                \null\hfill UMPP \#94--167}}
\title{Constraints on the finite-density spectral densities \linebreak of
vector channel}
\author{Xuemin Jin}
\address{Department of Physics and Center for Theoretical Physics\\
University of Maryland, College Park, Maryland 20742}
\date{\today}
\maketitle
\begin{abstract}
Sum rules for the variation of finite-density spectral density of
vector channel with baryon density are derived based on dispersion
relations and the operator product expansion. These sum rules may serve
as constraints on the phenomenological models for the finite-density
spectral densities used in the approaches motivated from QCD. Applying
these sum rules to the rho meson in nuclear medium with a simple
pole-plus-continuum ansatz for the spectral densities, we found that
the qualitative features of the QCD sum-rule predictions for the
spectral parameters are consistent with these sum rules; however, the
quantitative QCD sum-rule results violate the sum rules to certain
degree.
\end{abstract}

Study of hadronic properties at finite baryon density and temperature
is essential in understanding the structure of matter. Recently, this
subject has attracted much attention, motivated by the experimental
attainment of hot and dense matter in heavy-ion collisions and by the
theoretical expectations of a phase transition of matter from a
hadronic phase to a quark-gluon plasma at high temperatures.

Despite the difficulties due to the nonperturbative features of QCD at
large distances, one may study the properties of hadrons and the QCD
vacuum by investigating the two-point correlation functions of
currents, carrying the quantum numbers of the system under study. This
approach is based on the analytic properties of the two-point correlation
functions and on asymptotic freedom.  The hadronic spectral
properties ({\it e.g.,} masses, coupling constants, etc.) appearing in
the spectral densities can be related via dispersion relations to the
correlation functions evaluated in terms of quark and gluon degrees of
freedom.  In practical applications such as QCD sum-rule
method\cite{shifman1,reinders1} and analyses of lattice QCD
data\cite{chu1} and interacting instanton model
calculation\cite{shuryak1}, one needs to parametrize the spectral
functions with a small number of parameters and to evaluate the
correlation functions approximately [{\it e.g.,} operator product
expansion (OPE), lattice simulations, and interacting instanton
approximation]. The success of such approaches depends on correct
understanding of the qualitative features of the spectral functions and
accurate evaluation of the correlation functions from QCD.

In this paper, we derive sum rules for the variation of finite-density
spectral densities of vector channel with baryon density on the basis
of dispersion relation and the OPE. These sum rules can be regarded as
constraints on the phenomenological models for the finite-density
spectral densities. We also apply these sum rules to the rho meson in
nuclear medium with a simple pole-plus-continuum model for the spectral
densities. We find that the qualitative features of the QCD sum-rule
predictions for the spectral parameters are consistent with these
constraints; the quantitative QCD sum-rule results, however, violate
the constraints to certain degree.

Let us start with the correlation function of vector current at
finite baryon density and zero temperature\cite{hatsuda1,asakawa}:
\begin{equation}
\Pi_{\mu\nu}(q,\rho_B)\equiv i\int d^4x e^{i q\cdot x}\langle {\rm T}
J_\mu(x)J_\nu(0)\rangle_{\rho_B}\ ,
\label{corr-fd}
\end{equation}
where $J_\mu(x)$ is a vector current constructed from light quark
fields [{\it e.g.}, $J_\mu(x)={1\over 2}\left(\overline{u}(x)\gamma_\mu
u(x)\pm\overline{d}(x)\gamma_\mu d(x)\right)$]. Throughout this paper,
the up and down quark masses are taken to be equal and all the quark
fields in the current are assumed to have the same mass.  The notation
$\langle\cdots\rangle_{\rho_B}$ denotes the expectation value on the
finite-density ground state characterized by the baryon density
$\rho_B$ in the rest frame and the four-velocity $u^\mu$.
In medium, there are in general two independent invariants in the
vector channel, corresponding to the transverse and longitudinal
polarizations $\Pi_t(q,\rho_B)$ and $\Pi_l(q,\rho_B)$. For simplicity,
we will work in the rest frame of the medium, where $u^\mu=(1,{\bf
0})$, and take the three momentum to be zero ${\bf q}=0$. Then, since
there is no specific spatial direction, the transverse polarization is
related to the longitudinal one,
$\Pi_t(q_0,\rho_B)=q_0^2\Pi_l(q_0,\rho_B)$, where
$\Pi_l(q_0,\rho_B)=\Pi^\mu_\mu(q_0,\rho_B)/(-3q_0^2)$\cite{bochkarev1,hatsuda1,asakawa}.
The longitudinal part, $\Pi_l(q_0,\rho_B)=\Pi_l(q_0^2,\rho_B)$,
satisfies the standard dispersion relation\cite{galitskii,hatsuda1}
\begin{equation}
\Pi_l(q_0^2,\rho_B)=\int^\infty_0 ds {\rho(s,\rho_B)\over s-q_0^2}\ ,
\label{dis-fd}
\end{equation}
where $\rho(s,\rho_B)=\pi^{-1}{\rm Im} \Pi_l(s,\rho_B)$ is the
finite-density spectral density. Here we have omitted the subtraction
terms, which can be eliminated by taking derivatives of both sides of
Eq.~(\ref{dis-fd}) with respect to $q_0^2$. The sum rules to be derived
are, however, independent of this process.

Using Eq.~(\ref{dis-fd}), one can write the difference of the correlation
functions evaluated at different baryon densities as
\begin{equation}
\Delta\Pi_l(Q^2)\equiv\Pi_l(Q^2,\rho_B)-\Pi_l(Q^2,\rho_B^\prime)=\int^\infty_0
ds {
\rho(s,\rho_B)-\rho(s,\rho_B^\prime)\over s+Q^2}\ ,
\label{dis-fd-dif}
\end{equation}
where $Q^2\equiv -q_0^2$ and $\rho_B^\prime$ denotes a different baryon density
from $\rho_B$.

At very short distances, or at very high energies, the difference
between two correlation functions with different densities should go to
zero due to asymptotic freedom of QCD.  We can now look for the
consequences of this statement for the difference of spectral
densities. Expanding the right-hand side of Eq.~(\ref{dis-fd-dif}) for
large values of $Q^2$ we get
\begin{equation}
\Delta\Pi_l(Q^2)=\int^\infty_0 ds
\left[\rho(s,\rho_B)-\rho(s,\rho_B^\prime)\right]\left[
{1\over Q^2}-{s\over Q^4}+{s^2\over Q^6}-\cdots\right]\ .
\label{exp-fd-dif}
\end{equation}

On the other hand, for large $Q^2$ (i.e., in the deep Euclidean
region), one can evaluate the correlation functions by expanding the
product of currents according to the operator product expansion, which
leads to
\begin{equation}
\Pi_l(Q^2,\rho_B)=\sum_{n}C_n(Q^2)\langle\widehat{O}_n\rangle_{\rho_B}\ ,
\label{ope-gen}
\end{equation}
where $C_n(Q^2)$ are the Wilson coefficients and $\widehat O_n$ are
local composite operators constructed from quark and gluon fields. Here
we have suppressed the dependence of both the coefficients and the
operators on the normalization point $\mu$. The operators $\widehat
O_n$ are ordered by dimension (measured as a power of mass) and the
$C_n(Q^2)$ for higher-dimensional operators fall off by corresponding
powers of $Q^2$. The Wilson coefficients only depend on QCD Lagrangian
parameters such as the quark masses and the strong coupling constant;
{\it all density dependence} of the correlation function is included in
the condensates $\langle\widehat{O}_n\rangle_{\rho_B}$\cite{hatsuda1}.
Thus, one can express $\Delta\Pi_l(Q^2)$ as
\begin{equation}
\Delta\Pi_l(Q^2)=\sum_{n}C_n(Q^2) \Delta\langle\widehat{O}_n\rangle\ ,
\label{ope-dif}
\end{equation}
where
\begin{equation}
\Delta\langle\widehat{O}_n\rangle\equiv \langle\widehat{O}_n\rangle_{\rho_B}-
\langle\widehat{O}_n\rangle_{\rho_B^\prime}\ .
\end{equation}

Note that the pure perturbative contribution [corresponding to the unit
operator term, $\widehat{O}_n={\bf 1}$, in the OPE] to the correlation
function is independent of density, and thus does not appear in the
difference $\Delta\Pi_l(Q^2)$. The lowest order contribution to
$\Delta\Pi_l(Q^2)$ then comes from the condensates with lowest
dimension ($\widehat{O}_n\neq {\bf 1}$), which, in the vector channel,
are dimension four (including quark masses) condensates.  Since
$\Delta\Pi_l(Q^2)$ has dimension zero, the lowest order term in the OPE
of $\Delta\Pi_l(Q^2)$ must be proportional to $1/Q^4$. However, the
lowest order term in the phenomenological representation
Eq.~(\ref{exp-fd-dif}) is proportional to $1/Q^2$. Therefore, we
conclude that
\begin{equation}
\int_0^\infty\left[\rho(s,\rho_B)-\rho(s,\rho_B^\prime)\right]ds=0\ .
\label{sum-0}
\end{equation}
This is a rigorous result. In the OPE framework it simply follows from
the observation that the pure perturbative contribution is density
blind and the lowest-dimensional condensates have dimension four.
Although the finite-density spectral densities cannot be measured
directly from experiments, Eq.~(\ref{sum-0}) will constrain the change
of the spectral density with baryon density. Phenomenological
parametrizations often used in applications such as QCD sum-rule
calculations or analyses of lattice QCD data, must satisfy this
constraint.  If one adopts a pole-plus-continuum ansatz for the
spectral density,  Eq.~(\ref{sum-0}) indicates that the change in the
coupling for the pole is equal to the shift in the continuum.

The OPE of $\Pi_l(Q^2,\rho_B)$ takes the general form
\begin{equation}
\Pi_l(Q^2)=\sum_i c_i^{(4)} {\langle\widehat{O}_i^{(4)}\rangle_{\rho_B}
\over Q^4}
+\sum_j c_j^{(6)}{\langle\widehat{O}_j^{(6)}\rangle_{\rho_B}\over Q^6}+\cdots\
,
\label{ope-gen-pil}
\end{equation}
where the ellipses denote the contributions of condensates with higher
dimensions, the superscript indicates the dimension of operators, and
the sum is over all contributing operators with a given dimension. The
coefficients $c_l^{(d)}$ are dimensionless, and can, in principle, be
dependent on $Q^2$ due to the perturbative corrections [only through
strong coupling constant $\alpha_s(Q^2)$ and
$m_q^2/Q^2$]\cite{shifman1}. For simplicity, we will work only to the
lowest order in the strong coupling constant and to the first order in
quark masses, where the coefficients $c_l^{(d)}$ become independent of
$Q^2$.

We can then rewrite Eq.~(\ref{ope-dif})
as
\begin{equation}
\Delta\Pi_l(Q^2)=\sum_i c_i^{(4)} {\Delta \langle\widehat{O}_i^{(4)}\rangle
\over Q^4}
+\sum_j c_j^{(6)}{\Delta \langle\widehat{O}_j^{(6)}\rangle\over Q^6}+\cdots\ .
\label{left-hand-gen}
\end{equation}
Comparing the coefficients of $1/Q^4$ in Eqs.~(\ref{exp-fd-dif}) and
(\ref{left-hand-gen}), one obtains
\begin{equation}
\int_0^\infty s\left[\rho(s,\rho_B)-\rho(s,\rho_B^\prime)\right]ds=
-\sum_i c_i^{(4)} \Delta \langle\widehat{O}_i^{(4)}\rangle\ .
\label{sum-1}
\end{equation}
Similarly, equating the coefficients of $1/Q^6$, one finds
\begin{equation}
\int_0^\infty s^2\left[\rho(s,\rho_B)-\rho(s,\rho_B^\prime)\right]ds=
\sum_i c_i^{(6)}\Delta \langle\widehat{O}_i^{(6)}\rangle\ .
\label{sum-2}
\end{equation}

Following the same pattern, one may derive an infinite series of sum
rules, one for each OPE term (with fixed dimension) at small
distances.  The variation of the spectral density with baryon density
must satisfy these sum rules. If one takes $\rho_B^\prime=0$ (i.e., in
vacuum), these sum rules will constrain the change of the
finite-density spectral density relative to the corresponding vacuum
spectral density, which in some cases are experimentally accessible.
In the QCD sum-rule applications or the analyses of lattice simulation
data, one often parametrizes the spectral densities with a pole
representing the lowest resonance plus continuum contribution roughly
approximated by a perturbative evaluation of the correlation function,
starting at an effective continuum threshold. One may apply the above
sum rules to test this simple parametrization.

In principle, the sum rules described here can also be used to
determine the finite-density spectral density provided that its
corresponding vacuum spectral density and the values of in-medium and
vacuum condensates are known. In practice, however, one has to truncate
the OPE as the number of condensates with the same dimension appearing
in higher order terms become larger and there is no reliable way to
estimate these condensates. As a result, one can only expect to utilize
the first few sum rules (at best), which, along with a simple ansatz
for the spectral density, may give rise to an estimate of the
finite-density spectral parameters.

We notice, however, that the sum rules of higher order (resulting from
higher powers of $1/Q^2$) are more sensitive to the difference of the
spectral density in higher energy region. In the QCD sum-rule approach,
the Borel transformation suppresses the contribution from the higher
energy region (continuum), though it introduces an auxiliary parameter
(i.e., the Borel mass). Consequently, the spectral integral is
saturated by the lowest resonance; the roughness of the approximation
for the continuum is expected to have only small impact on the spectral
parameters for the lowest resonance\cite{shifman1}.  The sum rules
discussed here do not depend on any auxiliary parameter. However, for
the sum rules to be useful in determining the spectral properties of
hadrons, one needs to have a reliable model for the continuum.

The sum rules derived in the present paper may look like the usual
finite energy sum rules\cite{hatsuda1,krasnikov1,narison1}.
However, we emphasize that in our study, we focuses on the {\it
difference} of the correlation function evaluated at two different
baryon densities,  instead of the correlation function at a particular
density. The present sum rules are for the {\it variation} of the
finite-density spectral densities with baryon density, instead of the
spectral density at a particular density. Our derivation relies on the
subtraction procedure and on the asymptotic freedom of QCD, which
allows a short distance expansion of both the correlation function and
the phenomenological dispersion relation. This technique has been used
by Kapusta and Shuryak\cite{kapusta1} in deriving the Weinberg-type sum
rules at zero and finite temperature.  One may derive analogous sum
rules by using other type of subtraction scheme, instead of different
baryon density (see for example Ref.~\cite{huang1}). One can also
extend the sum rules to other channels, which will be documented in
Ref.~\cite{jinf}.

We now turn to apply the sum rules Eqs.~(\ref{sum-0}), (\ref{sum-1})
and (\ref{sum-2}) to the rho meson in nuclear medium, where the vector
current  is given by
$J_\mu(x)=\case{1}{2}\left[\overline{u}(x)\gamma_\mu
u(x)-\overline{d}(x)\gamma_\mu d(x)\right]$\cite{shifman1}. Since the complete
spectral density in nuclear medium is not known experimentally, one
cannot test directly whether the sum rules are indeed satisfied.

Various investigators\cite{hatsuda1,asakawa} have studied the
properties of the rho meson in nuclear medium within QCD sum-rule
approach.  It is found that the rho meson mass (pole position of the
propagator), the coupling of the vector current to the rho meson, and
the effective continuum threshold all drop in nuclear medium. At
nuclear matter saturation density, the rho meson mass decreases by
$\sim 15-18\%$ relative to its vacuum value. One can test whether these
in-medium spectral features predicted from QCD sum-rule calculations
are consistent with the sum rules derived in the present paper.

The explicit OPE result for the correlation function can be found in
Refs.~\cite{hatsuda1,asakawa}. To the first order in the nucleon
density $\rho_N$, the in-medium condensates can be written as
$\langle\hat{O}\rangle_{\rho_N}=\langle\hat{O}\rangle_{\rm
vac}+\langle\hat{O}\rangle_N\rho_N$, where $\langle\hat{O}\rangle_N$ is
the spin averaged nucleon matrix element\cite{hatsuda1,jin1}. This
linear approximation to the condensates is expected to be reasonable up
to the nuclear matter saturation density\cite{cohen3,hatsuda1}.
 Up to dimension six and to the linear order in
$\rho_N$, the result for the difference $\Delta\Pi_l(Q^2)$ can be
written as
\begin{eqnarray}
\Delta\Pi_l(Q^2)&=&\Pi_l(Q^2,\rho_N)-\Pi_l(Q^2,\rho_N^\prime=0)=
{m_q\over Q^4}\langle\overline{q}q\rangle_N\rho_N+{1\over 24
Q^4}\langle{\alpha_s\over\pi}G_{\mu\nu}G^{\mu\nu}\rangle_N\rho_N
\nonumber
\\*[7.2pt]
& &+{M_N\over 4Q^4}A^{u+d}_2\rho_N
-{224\over 81 Q^6}\pi\alpha_s\langle\overline{q}q\rangle_{\rm
vac}\langle\overline{q}q\rangle_N\rho_N -{5M_N^3\over 24Q^6}A^{u+d}_4\rho_N\ ,
\label{ope-real}
\end{eqnarray}
where $m_q$ is the average of up and down quark masses, $M_N$ is the nucleon
mass and $A^{u+d}_n(\equiv A^u_n+A^d_n)$ is a moment of the parton
distributions in the deep inelastic scattering
%
$A^{q}_n(\mu^2)=2\int_0^1 dx
x^{n-1}\left[q(x,\mu^2)+(-1)^n\overline{q}(x,\mu^2)\right]$,
%
where $q(x,\mu^2)$ and $\overline{q}(x,\mu^2)$ are the scale-dependent
distribution functions for quarks and antiquarks (of flavor $q$) in the
nucleon. Here the in-medium factorization approximation has been used
for the four-quark condensates [i.e.,
$\langle\overline{q}q\rangle_{\rho_N}^2-\langle\overline{q}q\rangle_{\rm
vac}^2\simeq 2\langle\overline{q}q\rangle_{\rm
vac}\langle\overline{q}q\rangle_N\rho_N$]\cite{hatsuda1}. In
Eq.~(\ref{ope-real}), we only retained the terms included in
Refs.~\cite{hatsuda1,asakawa}.

Since the sum rules Eqs.~(\ref{sum-0}) and (\ref{sum-1}--\ref{sum-2})
involve integrations of the spectral densities with different powers of
$s$, it is likely important to incorporate the finite widths of the
resonances and to have a reliable model for the continuum. Here we will
adopt the pole-plus-continuum parametrization for both vacuum and
in-medium spectral density, which is widely used in the QCD sum-rule
calculations\cite{shifman1,hatsuda1,leinweber1}, and hence only expect
to test the qualitative features of the finite-density spectral
parameters. The detailed account of the finite width of the rho meson,
along with the incorporation of a more reliable continuum model, will
be given in Ref.~\cite{jinf}.

In the pole-plus-continuum approximation, one can write the vacuum spectral
density as\cite{shifman1}
\begin{equation}
\rho_{\rm vac}(s)=F\delta(s-m_\rho^2)+{1\over 8\pi^2}\left(1+{\alpha_s\over
\pi}\right)\theta(s-s_0)\ ,
\label{anz-vac}
\end{equation}
and the finite-density spectral density as\cite{hatsuda1}
\begin{equation}
\rho(s,\rho_N)=\rho_{\rm sc}\delta(s)+F^*\delta(s-m_\rho^{*^2})+{1\over
8\pi^2}\left(1+{\alpha_s\over \pi}\right)\theta(s-s_0^*)\ ,
\label{anz-med}
\end{equation}
where $F$ ($F^*$), $m_\rho$ ($m_\rho^*$) and $s_0$ ($s_0^*$) are the
coupling, the rho meson mass and the continuum threshold, respectively,
and $\rho_{\rm sc}$ denotes the contribution of Landau damping (or the
scattering term)\cite{bochkarev1,hatsuda1}.

Substituting the spectral ansatz
Eqs.~(\ref{anz-vac}--\ref{anz-med}) and the OPE results into Eqs.~(\ref{sum-0})
and (\ref{sum-1}--\ref{sum-2}),
we obtain the following sum rules
\begin{eqnarray}
& &
\rho_{\rm sc}+F^*-F+{1\over 8\pi^2}\left[1+{\alpha_s\over
\pi}\right]\left(s_0-s_0^*\right)=0\ ,
\label{sum-real-0}
\\*[7.2pt]
& &F^*m_\rho^{*^2}-Fm_\rho^2+{1\over 16\pi^2}\left[1+{\alpha_s\over
\pi}\right]\left(s_0^2-s_0^{*^2}\right)=-m_q\langle\overline{q}q\rangle_N\rho_N
\nonumber
\\*[7.2pt]
& &\hspace{3in}
-{1\over
24}\langle{\alpha_s\over\pi}G_{\mu\nu}G^{\mu\nu}\rangle_N\rho_N-{M_N\over
4}A^{u+d}_2\rho_N \ ,
\label{sum-real-1}
\\*[7.2pt]
& &F^*m_\rho^{*^4}-Fm_\rho^4+{1\over 24\pi^2}\left[1+{\alpha_s\over
\pi}\right]\left(s_0^3-s_0^{*^3}\right)=
\nonumber
\\*[7.2pt]
& &\hspace{2in}
-{224\over 81}\pi\alpha_s \langle\overline{q}q\rangle_{\rm
vac}\langle\overline{q}q\rangle_N\rho_N-{5M_N^3\over 24}A^{u+d}_4\rho_N \ ,
\label{sum-real-2}
\end{eqnarray}
where we have assumed that the in-medium continuum threshold $s_0^*$ is less
than its vacuum value $s_0$. Substituting the QCD sum-rule predictions for the
spectral parameters into these sum rules one may check how well these sum rules
are satisfied. Alternatively, one may extract the in-medium spectral parameters
by solving these three equations and compare the results with the QCD sum-rule
predictions. Here we follow the latter.

To obtain the finite-density spectral parameters $m_{\rho^*}$, $F^*$,
and $s_0^*$ from Eqs.~(\ref{sum-real-0}--\ref{sum-real-2}), one needs
to know the various nucleon matrix elements appearing on the right hand
sides as well as the corresponding vacuum spectral parameters. The
nucleon matrix element $\langle\overline{q}q\rangle_N$ is related to
the nucleon sigma term $\langle\overline{q}q\rangle_N=\sigma_N/2m_q$;
we take $\sigma_N\simeq 45\,\text{MeV}$\cite{gasser1} and $m_q\simeq
5.5\,\text{MeV}$\cite{cohen3,jin1}.  For the gluon matrix element, we
use $\langle(\alpha_s/\pi)G_{\mu\nu}G^{\mu\nu}\rangle_N\simeq
-650\,\text{MeV}$\cite{cohen3,jin1}.
The moments of parton distribution are taken to be $A_2^{u+d}\simeq
0.938$ and $A_4^{u+d}\simeq 0.121$ (at
$\mu^2=1\,\text{GeV}^2$)\cite{asakawa}. We adopt
$\langle\overline{q}q\rangle_{\rm vac}\simeq
(-245\,\text{MeV})^3$\cite{furnstahl1} and $\alpha_s\simeq
0.3$\cite{asakawa} in our calculations. The nuclear matter saturation
density is taken to be $\rho_0=(110\,\text{MeV})^3$.  We fix
$m_\rho=770\,\text{MeV}$ and $s_0=1.5\,\text{GeV}^2$\cite{shifman1},
and parametrize the scattering term as $\rho_{\rm sc}=a_0\rho_N$.

In Fig.~\ref{fig-1}, the resulting ratio of the in-medium rho meson
mass to its vacuum value is plotted as function of nucleon density for
different values of $a_0$ and a fixed $F$ value $F=2 f_\pi^2$ with
$f_\pi=93.5\,\text{MeV}$, which is obtained by using
$F=m_\rho^2/g_\rho^2$\cite{sakurai1} with the KSFR relation
$g_\rho^2=m_\rho^2/2f_\pi^2$.  One notices that the in-medium rho meson
mass drops relative to its vacuum value.  Similar behavior is also
found for both the coupling and the continuum threshold. These
qualitative features are consistent with those predicted from the QCD
sum-rule calculations.

On the other hand, we observe that the quantitative result for the
ratio is sensitive to the value of $a_0$ (i.e., $\rho_{\rm sc}$). For
$a_0=0$, the in-medium rho meson mass is only few percent smaller than
its free space value even at the nuclear matter density. For
$a_0=3.6\,\text{GeV}^{-1}$, the rho meson mass drops to $\sim 0.8
m_\rho$ at the saturation density. In Ref.~\cite{asakawa} the
scattering term is neglected (i.e., $a_0=0$) while in
Ref.~\cite{hatsuda1} it is taken to be $a_0=1/2M_N$.  The ratio
$m_\rho^*/m_\rho$ is found to be $\sim 15-18\%$ at $\rho_N=\rho_0$ in
these references.
However, we find that to obtain a $15-18\%$ decrease in the rho mass at
the nuclear matter saturation density, one needs to use a value of
$a_0\sim 3\,\text{GeV}^{-1}$, which is much larger than that used in
Ref.~\cite{hatsuda1}. This inconsistency signals that the simple
pole-plus-continuum model for the spectral densities {\it with} the QCD
sum-rule predictions for the spectral parameters violates the
constraints Eqs.~(\ref{sum-0}) and (\ref{sum-1}--\ref{sum-2}) to
certain degree. A improved model beyond this simple parametrization for
the spectral densities may be needed to satisfy the constraints.

In Fig.~\ref{fig-2}, we plot the ratio $m_\rho^*/m_\rho$ as a function
of the nucleon density for different values of $F$ with fixed
$a_0=1/2M_N$. It can be seen that the ratio is very sensitive to $F$,
in particular for small $F$ values. (For $F\ge 0.035\,\text{GeV}^2$,
there is no real solution). Again, we note that a much smaller value of
$F$ is necessary to reproduce the QCD sum-rule result. It is also found
that our result is relatively insensitive to the values of $m_\rho$ and
$s_0$.

In conclusion, we have derived sum rules for the variation of the
finite-density spectral density of vector channel with baryon density
within the framework of operator product expansion and the dispersion
relation. These sum rules may serve as constraints on the
phenomenological models used in the QCD sum-rule calculations or in the
interpretation of the lattice QCD data.
We also noted that in principle one can use these sum rules to
determine the qualitative properties of the finite-density spectral
parameters if the corresponding vacuum spectral density and the
in-medium and vacuum condensates are known.  We applied the first three
sum rules to the rho meson in nuclear medium with a simple
pole-plus-continuum parametrization for the spectral densities, and
found that the qualitative features of the QCD sum-rule predictions are
consistent with our sum rules. However, the quantitative result shows
that the simple ansatz with QCD sum-rule predictions for the spectral
parameters violates the sum rules to certain degree.

This suggests that the inclusion of the finite widths of the
resonances and the refinement of the continuum model may be important.
This point, along with the full detail of the present paper and its
extension to other channels, will be reported elsewhere\cite{jinf}.

\vspace{0.5in}

The author would like to thank G. E. Brown, M. K. Banerjee, T. D.
Cohen, H. Forkel, and D. K. Griegel for stimulating discussions and
helpful comments. This work is supported by the U.S. Department of
Energy under grant No. DE-FG02-93ER-40762 and U.S. National Science
Foundations under grant No.  PHY-9058487.

\begin{figure}
\caption{Ratio $m_\rho^*/m_\rho$, obtained  from solving
Eqs.~(\protect{\ref{sum-real-0}}--\protect{\ref{sum-real-2}}), as a function of
the nucleon density.
The four curves correspond to $a_0=0$ (solid), $a_0=1.2\,\text{GeV}^{-1}$
(long-dashed), $a_0=2.4\,\text{GeV}^{-1}$ (dotted), and
$a_0=3.6\,\text{GeV}^{-1}$ (dashed). The other input parameters are
described in the text.}
\label{fig-1}
\end{figure}
\begin{figure}
\caption{Ratio $m_\rho^*/m_\rho$, obtained from solving
Eqs.~(\protect{\ref{sum-real-0}}--\protect{\ref{sum-real-2}}), as a function of
the nucleon density.
The four curves correspond to $F=0.005\,\text{GeV}^2$ (solid),
$F=0.01\,\text{GeV}^2$ (long-dashed), $F=0.02\,\text{GeV}^2$ (dotted),
and $F=0.03\,\text{GeV}^2$ (dashed). The other input parameters are the same as
in Fig.~\protect{\ref{fig-1}}.}
\label{fig-2}
\end{figure}

\end{document}